\documentclass[pre,twocolumn,showpacs,epsfig]{revtex4}
\usepackage{graphicx}
\usepackage{dcolumn}
\usepackage{amssymb}
\usepackage{bm}
\usepackage{epsfig}
\usepackage{psfrag}

\def\3dots{\:\raisebox{-0.5ex}{$\stackrel{\textstyle.}{:}$}\:}
\def\beq{\begin{equation}}
\def\eeq{\end{equation}}
\def\bea{\begin{eqnarray}}
\def\eea{\end{eqnarray}}

\begin{document}

\title{Tunable Brownian Vortex at the Interface}

\author{Manas Khan and A. K. Sood}
\affiliation{Department of Physics, Indian Institute of Science, Bangalore - 560012, India}

\date{\today}
\begin{abstract}

A general kind of Brownian vortexes are demonstrated by applying an external nonconservative force field to a colloidal particle bound by a conservative optical trapping force at a liquid-air interface. As the liquid medium is translated at a constant velocity with the bead trapped at the interface, the drag force near the surface provide enough rotational component to bias the particle's thermal fluctuations in a circulatory motion. The interplay between the thermal fluctuations and the advection of the bead in constituting the vortex motions is studied, inferring that the angular velocity of the circulatory motion offers a comparative measure of the interface fluctuations.           

\end{abstract}

\maketitle


Since the development of single beam gradient-force optical trap \cite{ashkin} it has become a very effective and useful tool to perform single-molecule experiments. Starting from the manipulation of nanoparticles to the study of cellular processes, optical tweezers are the most widely used mechanism to apply or measure forces as small as piconewton or even femtonewton \cite{rev}. In most of these experiments the optical tweezer is modeled as a Hookean spring, creating a harmonic potential well for the trapped particle. Though the conservative part of the force field, namely the gradient force can be described best in that manner, the scattering force which is nonconservative in nature is completely ignored. The influence of the scattering force field on the dynamics of a colloidal particle in an optical trap has been studied rather recently \cite{Grier1}. The solenoidal force field biases the thermal fluctuations of the trapped particle and drives it to a steady state where its probability flux traces out a toroidal vortex. This phenomenon, with a more detailed examination, has later been defined as Brownian vortex \cite{Grier2, Grier3}. However, this toroidal bias in the short-time motion of a trapped bead is too subtle to be perceived and do not affect the experiments in any way, unless, (a) the power of the trapping beam is extremely low to permit the trapped particle to access a large volume, and (b) a very long trajectory or the ensemble average of many short trajectories is analyzed. With only a few studies reported on this so far \cite{Grier1, Grier2, Grier3, volpeEPL, prl, nonspherical}, Brownian vortex remains as a very intriguing but comparatively less explored area in statistical physics.

A Brownian vortex is distinguished as to reflect the interplay between advection, under the influence of the nonconservative force field and diffusion of a particle about the point of minimum potential set by the conservative force field. Therefore it has commonly been observed in the case of a particle in a weak optical trap where the scattering force adds the nonconservative part to an otherwise conservative force field generated by the gradient force. However, since the scattering force and the gradient force are interconnected by many factors, e.g. the size of the particle, its refractive index contrast at the working wavelength etc., it is not possible to tune these two forces independently and thus to modulate the interplay. In this communication, we report a more general form of Brownian vortex that is operational even in strong optical traps with evidently visible circulatory bias in the thermal fluctuations of a trapped bead and thereby address some of its generic characteristics, which can also be used to probe the statistical properties of the medium. In our experiments we have applied an external solenoidal force field on a particle held in the potential well of an optical trap to constitute a Brownian vortex. This mechanism enables us to vary both the conservative and the nonconservative force fields independently and to study this phenomenon extensively in a wide domain.

We have trapped a colloidal bead at a liquid-air interface and translated the sample stage at a constant velocity in order to observe a Brownian vortex. As the sample stage is driven with a constant velocity, the drag force on the trapped particle at the interface provides a nonconservative force field that can be modulated easily by changing the speed of the drive. It is important to note that the lines of force created by the Stokes drag in this scenario are parallel to each other and equal in magnitude in the bulk. In other words, the Stokes drag experienced by a submerged bead while the surrounding fluid is moving at a constant velocity is independent of its position. Therefore a similar experiment performed with the bead trapped in the bulk liquid medium would not exhibit a Brownian vortex \cite{microrheology}. However, the viscosity and thereby the Stokes drag force changes dramatically as the liquid-air interface is approached \cite{sevick, bickel}. A particle experiences increased mobility (parallel to the interface), or reduced friction as it comes closer to the surface reflecting the interface's inability to support a shear stress. Near the liquid-air interface, which permits perfect slip, the viscosity becomes an increasingly sensitive function of the $z$-depth. In addition, the surface tension starts playing a role when the bead, trapped at the interface, is dragged along the surface. With all these factors the drag force field does not remain uniform near the interface and provide finite rotational component to bias the particle's thermal fluctuations in a circular motion. The simultaneous presence of advection of the particle due to the Stokes drag, which is a function of its $z$-position, and the Brownian diffusion in the optical trap results in a Brownian vortex.

\begin{figure}[htb]
\includegraphics[width=0.48\textwidth]{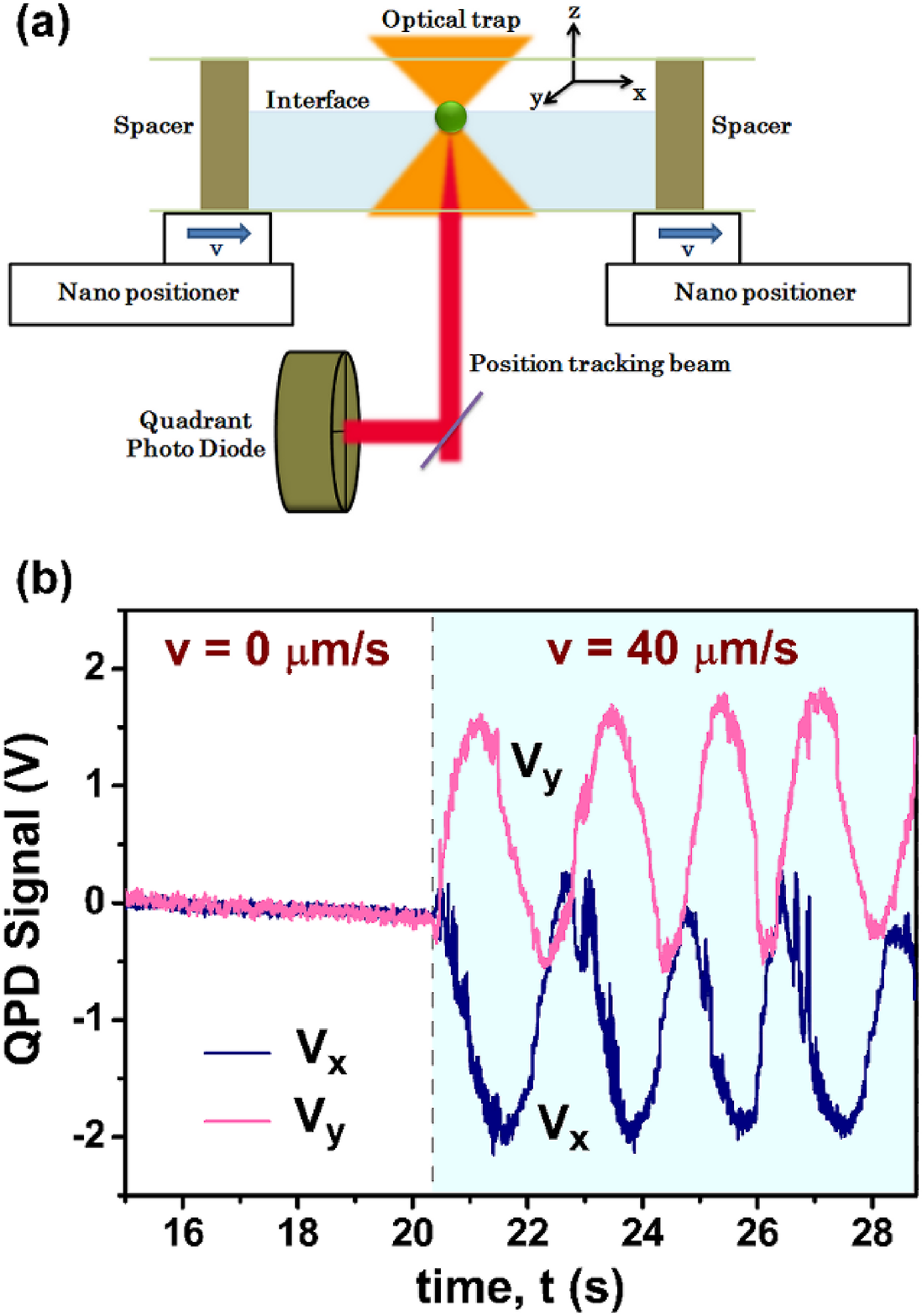}\\
\caption{(Color online) (a) The schematic of the experimental setup is shown. A colloidal bead is trapped at the liquid-air interface and its instantaneous position (in the $XY$-plane) is probed with the help of a quadrant photo diode (QPD). The sample cell is translated with a constant velocity $v$ while the bead is held stationary by the optical trap. (b) The QPD signals in both stage velocity off and on state are displayed. As the stage velocity is set on, the QPD signals show large sinusoidal variations where a strong correlation is apparent.} \label{setup}
\end{figure}

In our experiments a polystyrene bead of diameter 2.03 $\mu m$ is trapped at a water-air interface. The optical tweezer is formed by tightly focusing an infra-red ($\lambda$ = 1064 $nm$) laser beam through a 63$\times$, 1.4 $N.A.$ objective fitted to an inverted microscope. A sample cell is formed by two cover glasses with an O-ring sealed in between as spacer. A small amount of diluted colloidal aqueous suspension is loaded in the cavity of the O-ring before covering it with the top cover glass. The volume of the sample used is measured appropriately to make a 20-25 $\mu m$ high liquid column inside the sample cell. A colloidal bead is first trapped and then brought to the liquid-air interface which is fairly above the bottom plate so that any solid-liquid interface effect can be neglected in this experiment. The trapped bead is pulled up slowly to the interface by continuously moving the focal plane upwards until the trapped bead just goes out of focus. At this point the focal plane crosses the interface but the particle remains bound at the interface. The focal plane is then moved down a bit to bring back the trapped bead fully in focus. To track the particle's fluctuations a red ($\lambda$ = 680 $nm$) laser beam is sent along with the infrared laser beam. The trapped bead is imaged by this tracking laser beam on to a quadrant photo diode (QPD). A schematic of the experimental setup is shown in Fig \ref{setup}(a). The output current signals of the QPD (at 3 $kHz$) is amplified and converted into voltage signals that directly correspond to the particle's position fluctuations at the focal plane through the calibration factors. At this point the QPD signals reflect the diffusion motion of the trapped bead at the interface. The sample stage is then given a constant velocity $v$ along $\hat{x}$ by sending proper voltage signals to the nanopositioner which holds the sample cell. Every positive $X$ motion (positive value of $v$)  of 10$s$ duration is traced back with a similar negative $X$ motion (negative value of $v$) with a pause of 20$s$ in between and this cycle is iterated several times to improve the statistics of our data. The QPD signal is recorded continuously in this process. The experiment is repeated for four different stage velocities, from 10 $\mu m/s$ to 40 $\mu m/s$ and for two different samples, water and 1 $wt\%$ CTAT (cetyltrimethylammonium tosylate), at varying laser powers. The laser powers used in these experiments are 50 $mW$ (for water), 75 $mW$ and 100 $mW$ (for 1 $wt\%$ CTAT) at the sample plane, which correspond to trap stiffness of 4.9 $pN/\mu m$, 7.3 $pN/\mu m$ and 9.7 $pN/\mu m$ respectively, as measured in bulk water for the same bead.

\begin{figure}[htbp]
\includegraphics[width=0.5\textwidth]{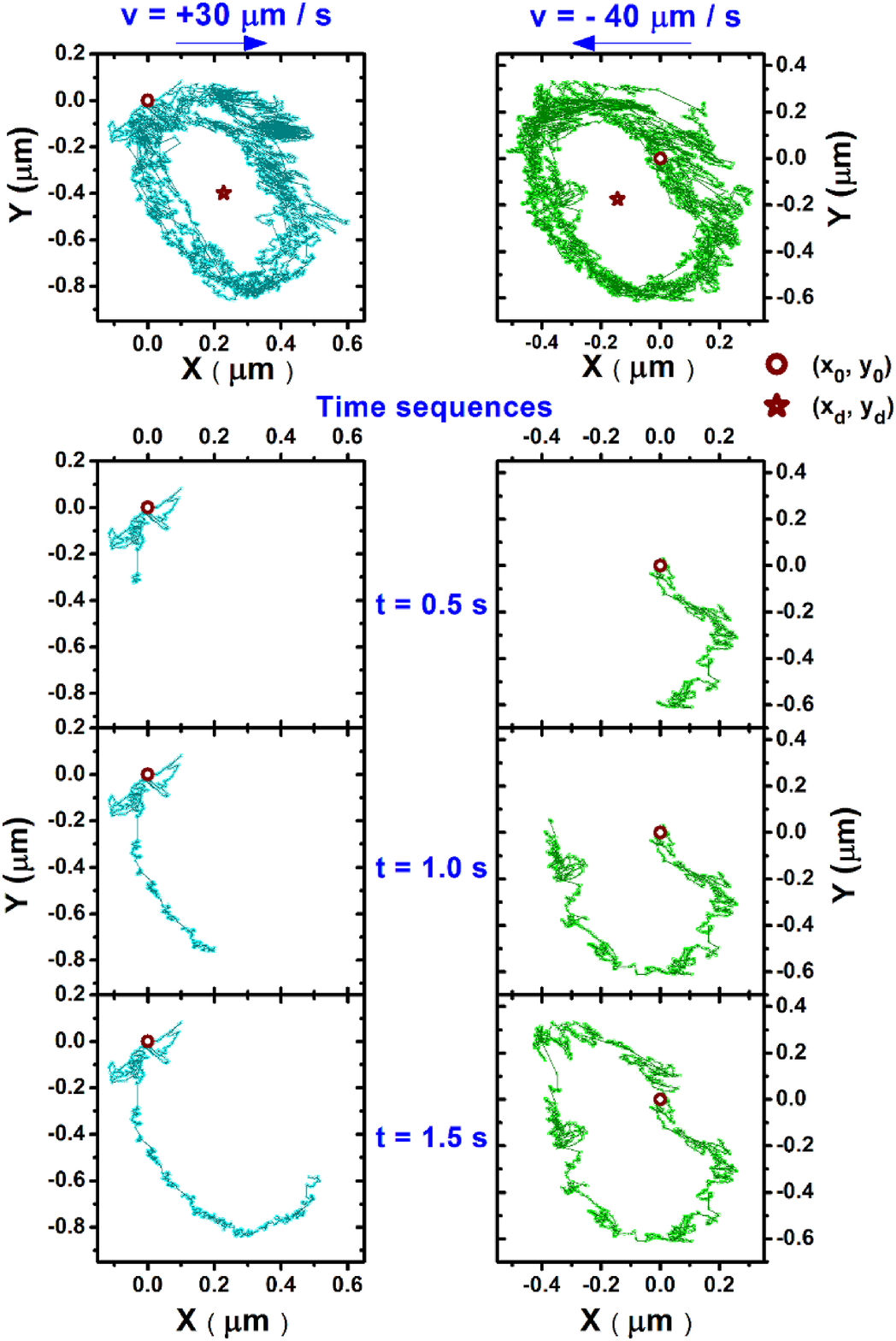}\\
\caption{(Color online) The phase diagram of two typical Brownian vortexes are displayed. In both the cases the colloidal particle is trapped at water-air interface. The sample stage velocities (along $X$) are $v$ = 30 $\mu m/s$ and - 40 $\mu m/s$ in the left and the right column respectively. The top row shows the circulatory motions for over a complete 10$s$, while the remaining rows exhibit three time snaps. The circle marks indicate the equilibrium trapping position, $(x_{0}, y_{0})$, which is the starting point of the observation; and the star marks denote the center of the vortexes, $(x_{d}, y_{d})$. Supplementary video files \cite{vid} show the animation of these vortex motions.} \label{rotation}
\end{figure}

As soon as the stage velocity ($v$) becomes non-zero the QPD signal changes dramatically, as shown in Fig. \ref{setup}(b). With $v \neq 0$, the QPD signals show a large sinusoidal variation in both $X$ and $Y$ positions and reflect a circulatory motion in the $XY$-plane when represented as a phase space plot (Fig. \ref{rotation}). This circulatory motion represents a 2D projection in the $XY$-plane of a 3D Brownian vortex. Because of the experimental limitations, the  instantaneous $z$-position of the bead cannot be probed simultaneously in this experiment. For these large modulations in $X$ and $Y$ signals the QPD sum signal (of all four quadrants) does not faithfully convey the position fluctuations along $Z$. In addition, the QPD sum signals from different experiments cannot be compared to comment on the initial $z$-position of the bead (when the variation in $X$ and $Y$ signals are not large) since the sum signal is sensitive to many other factors other than the $z$-position fluctuation of the bead and those factors change significantly from one experiment to another. Though the holographic imaging technique \cite{Grier1} allows simultaneous measurement of all the three coordinates of the bead's position, it was not employed because of its limitation in frame rate. In this study we preferred to capture the trapped bead's trajectory at high bandwidth (3000 data points per second) where the circular bias in the position fluctuations of the colloidal particle becomes visible. Nevertheless, the 2D projection that we observe here exhibits all the three distinguishing characteristics of a Brownian vortex \cite{Grier3}, as it (a) represents the diffusion of the particle in a static force field, (b) would attain a mechanical equilibrium (at a shifted $x$ position) in the absence of thermal fluctuations (along $Z$), and (c) has a steady-state probability flux. At the interface, the drag force becomes sensitive to the $z$-position of the bead and therefore the thermal fluctuations of the bead along $Z$ picks up a strong rotational bias to constitute a Brownian vortex that is visibly evident (Fig. \ref{rotation}). A similar experiment with the particle trapped in the bulk fluid does not manifest a Brownian vortex, rather probes the microrheological properties of the medium \cite{microrheology} because the force field produced by Stokes drag does not have non-zero rotational component there. It would be worth mentioning that in the present case both the gradient force field and the static nonconservative force field are quite strong compared to the available thermal energy of the trapped bead, which indicates that this Brownian vortex can work even when the thermal fluctuations are small.

Two typical circulatory motion of a trapped bead at the water-air interface for stage velocities $v$ = 30 $\mu m/s$ and $v$ = -40 $\mu m/s$ are displayed in Fig. \ref{rotation} at left and right column respectively. The first row exhibits the motion for entire 10$s$ duration and the remaining rows show the time snaps to represent the dynamical characteristics. It is apparent from the time snaps that the angular velocity ($\Omega$) increases with increasing stage velocity. However, the direction of the circulatory motion changes randomly in repeated observations at the same settings without having any correlation with the direction of stage velocity. This property could be attributed to the fact that the plane of the vortex motion in 3D can make a positive or negative angle, in an arbitrary fashion, with the $XZ$-plane thereby causing a change in the direction of the observed circulation which is a 2D projection on the $XY$ plane. The starting point of the vortex, which is also the equilibrium position of the trapped bead in drive off state, is denoted by ($x_{0},y_{0}$) and the center (average position) of the vortex is represented as ($x_{d},y_{d}$). While the displacement along $X$, $\Delta x$ ($= x_{d} - x_{0}$) shows a deterministic dependence on the direction of stage velocity $v$, $\Delta y$ ($= y_{d} - y_{0}$) takes positive or negative values in a completely random manner. The stochastic behavior of $\Delta y$ and the direction of the circular motion go well along with the above argument.

\begin{figure}[htb]
\includegraphics[width=0.47\textwidth]{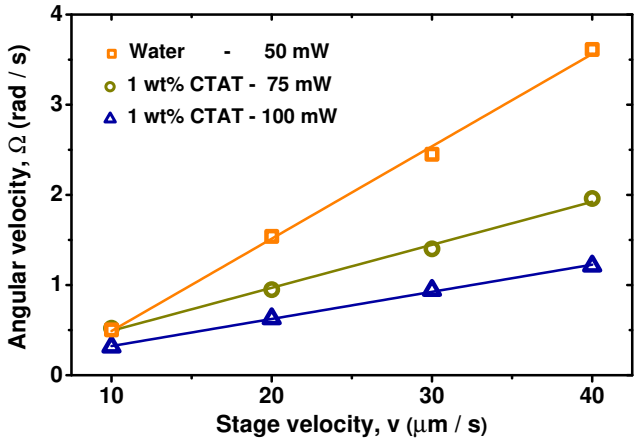}\\
\caption{(Color online) The angular velocity $\Omega$ of the Brownian vortexes have been plotted against the stage translation speed $v$ for three different cases. The lines are the linear fit to the plots.}
\label{speed}
\end{figure}

The circulatory motion of the trapped bead at the interface has been studied extensively at different settings. To investigate the influence of interface fluctuations on this Brownian vortex, the experiment has also been performed at the liquid-air interface of a viscoelastic fluid. The surfactant CTAT solution (1 $wt\%$) in water has higher surface tension which ensures lower level of interface fluctuations. Brownian vortexes are observed at the liquid-air interface for both the samples, though with different angular velocities. The experiments with 1 $wt\%$ CTAT solution were done at different laser powers. While the experiments at water-air interface were performed at 50 $mW$ laser power, 75 $mW$ and 100 $mW$ laser powers have been used for the case of 1 $wt\%$ CTAT, with the same stage velocities ($v$). A higher laser power, and hence, a stronger trap helps balancing the greater drag force in experiments with the CTAT sample. Fig. \ref{speed} displays the angular velocity ($\Omega$) of the ciculatory motions at varying stage velocities, $v$, for all the three cases. They exhibit a linearly increasing trend with $v$ in each case irrespective of the sample or the laser power used, demonstrating the influence of the advective force in the vortex motion. The angular velocity, $\Omega$, of the Brownian vortexes at the water-air interface increases most sharply with the stage velocity ($v$). In case of the CTAT-air interface, the angular velocities are less, and also less sensitive to the stage velocity, $v$, as compared to the water-air interface. It is also evident from the plots that a higher laser power results in a slower angular velocity of the vortex motion as the position fluctuations of the bead become more restricted. In other words, a stronger conservative force field confines the particle to a greater degree where the reduced Brownian fluctuations extract less amount of work from the nonconservative force field. This clearly manifests the origin of the observed circulatory motions of the trapped bead at the interface and recognizes them as the Brownian vortexes.

It would be important to clarify here that the observed circulatory motions of the trapped bead at the liquid-air interface are free from other effects i.e. thermal convection. Local heating does not cause a convective current here as before switching on the stage velocity the trapped bead does not show any circular motion. With increasing laser power the speed of rotation does not show an increment, it rather decreases.

\begin{figure}[htb]
\includegraphics[width=0.47\textwidth]{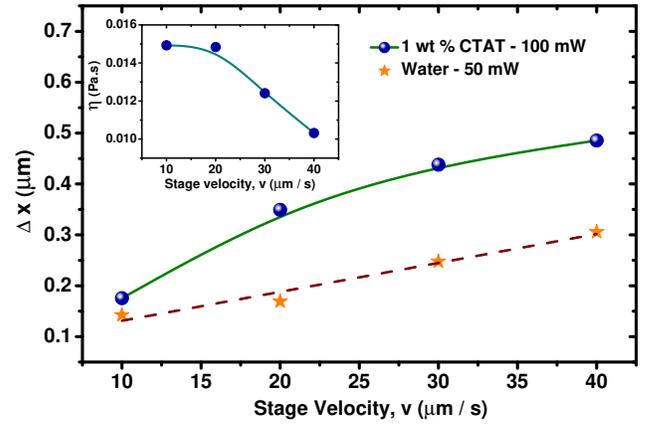}\\
\caption{(Color online) The displacement along $X$, $\Delta x$, has been plotted against the stage velocity $v$ for two different cases. The dashed line is a linear fit to the water data. The solid line is a guide to eye for the CTAT data. Inset: Shear thinning at CTAT-air interface. The solid line is a guide to eye.}
\label{displacement}
\end{figure}

A trapped particle inside a moving bulk fluid reaches to a mechanical equilibrium at a displaced position and its displacement $\delta x$ from the trap center compares the Stokes drag with the spring constant $\kappa$ of the optical trap through the equation $\kappa \, \delta x = 6 \pi \eta a v$, where the right hand side is the Stokes drag, $\eta$, $a$ and $v$ being the viscosity of the fluid, radius of the bead and the velocity of the fluid respectively. However, the Stokes drag changes significantly and dramatically as the liquid-air interface is approached, where even the value of $\eta$ gets modified from that in the bulk. The displacement of the trapped bead would still provide a comparison of the nonconservative drag force to the conservative gradient force. For these Brownian vortexes at the interface, the trapped particle maintains a dynamic equilibrium with a steady-state probability flux circulating around the point $(x_{d}, y_{d})$, which in other words represents the average position of the particle in the drive on-state. The displacements $\Delta x$ has been plotted against the stage velocities $v$ for two comparable cases in Fig. \ref{displacement}. A greater drag force at the CTAT-air interface is balanced by a stronger trap and results in a comparable displacement $\Delta x$. For the water-air interface, $\Delta x$ follows a linear trend with the stage velocity $v$, while shear thinning is observed at the CTAT-air interface. Applying the Stokes drag equation at the interface, viscosity $\eta \, (=\frac{\Delta x}{v}\frac{\kappa}{6 \pi a})$ is calculated, taking $\kappa$ value for the CTAT-air interface (determined from $\left\langle \delta x^{2} \right\rangle$ at $v = 0$), at varying stage velocities, as shown in the Inset of Fig. \ref{displacement}. The calculated viscosities in this case are about 15\% smaller that that obtained in an active microrheology experiment in the bulk  \cite{microrheology}. Though the trapped particle exhibit Brownian vortexes at the moving interface, its average position displacements nonetheless show similar trends as that of a bead trapped inside the bulk medium.

The ratio of the drag force to the gradient force, given by $\Delta x$, is quite comparable in the two cases as shown in Fig. \ref{displacement}. However, the angular velocity of the Brownian vortexes at those two settings, water-air interface at 50 $mW$ laser power and CTAT-air interface at 100 $mW$ laser power, are very different. This clearly establishes that a reduced $z$-position fluctuations of the bead at the CTAT-air interface, because of less interface fluctuations, in addition to a stronger trap, compared to the case of water-air interface, cannot extract as much energy even from an enhanced nonconservative force field. If these experiments are performed on two liquid samples with similar viscosity but different surface tensions, the influence of the interface fluctuations in formation of a Brownian vortex at the interface would be more prominent.

In summary, we have observed a general kind of Brownian vortex where an external nonconservative force field has been applied to bias the thermal fluctuations of a harmonically bound particle in a circulatory motion. The position fluctuations of a colloidal bead trapped at a liquid-air interface extract work from the rotational component of the drag force field near the surface to constitute a Brownian vortex. The angular velocity of this vortex motion is directly correlated to the particle's position fluctuations at the interface, which offers an useful probe to different statistical properties of the interface; interface fluctuations, or intermittency therein (may be caused by an external shear drive) and the frequency dependent surface tension, or in a more general sense, the viscoelastic properties of the interface being the most important and interesting examples.        

We thank Council for Scientific and Industrial Research (CSIR), India for financial support through Bhatnagar fellowship to AKS.

\end{document}